\documentclass[preprint,12pt,num]{elsarticle}

\usepackage{amssymb}
\usepackage{amsmath}
\usepackage{textcomp}
\usepackage[squaren]{SIunits}

\journal{ArXiv}

\begin{document}

\begin{frontmatter}

\title{Reproducibility of Large Eddy Simulations for mixing in stirred tank reactors} 

\author[inst1]{Cees Haringa\corref{cor1}}
\ead{c.haringa@tudelft.nl}

\author[inst2]{Ryan Rautenbach}
\author[inst1]{Héctor Maldonado de León}
\author[inst1]{Pieter Brorens}
\author[inst2]{Michael Schlüter}

\affiliation[inst1]{organization={Department of Biotechnology, Delft University of Technology},
            addressline={Van der Maasweg 9}, 
            city={Delft},
            postcode={2629 HZ},
            country={The Netherlands}}

\affiliation[inst2]{organization={Institute of Multiphase Flows, Hamburg University of Technology, Hamburg, Germany},
            addressline={Eißendorfer Straße 38}, 
            city={Hamburg},
            postcode={21073},
            country={Germany}}

\cortext[cor1]{Corresponding author}

\begin{abstract}
CFD simulations are widely used to quantify mixing performance of stirred tanks, for various applications in chemical engineering and biotechnology. Due to advances in GPU computing, more and more often these simulations make use of Large Eddy Simulations (LES), which explicitly simulate the dynamics of large-scale turbulence. Although these simulations are fully deterministic and hence theoretically reproducible, small numerical variations induced by round-off errors combined with differences in distribution and order of operations in parallel computing lead to separation of trajectories, i.e. different flowfield evolutions and different mixing times between repeat simulations, even on the same architecture. We investigate the impact of repeat simulations on the mixing time distribution observed in a $30 \liter$ stirred vessel with two commercial CFD packages, and compare to experimental variability. While the distribution between simulations and experiments is in very good agreement, we do conclude confidence intervals should be reported for CFD simulations of mixing. 
\end{abstract}




\end{frontmatter}

\section{Introduction}
\label{intro}
Mixing is a common operation in many industrial operations, and as such has been thoroughly researched over many decades, both experimentally \citep{kramers1953,nienow2014} and numerically \citep{ranade1991,akker2006,hartmann2006}. Many methods have been developed to experimentally quantify mixing, from invasive probe-based (single-point) approaches to detailed spatial methods based on pH indicators \citep{fitschen2021}. This makes it impossible to assign a single mixing time to a system, as both the means of measurement and the mixing criterion (e.g. 90 or 95$\%$ homogenization) vary; furthermore, prior work has shown that even mimicking identical setups between different facilities may lead to variability in the order of 15$\%$ \citep{kraume2001}. What is not surprising, however, is that a mixing time distribution is practically observed in turbulent stirred vessels (single phase); measurement errors aside, turbulent mixing is a stochastic process, that may additionally be impacted by macro-instabilities that can lead to strong mixing time variations \citep{ducci2007}. Still, many experimental studies regularly report a single mixing time, or an average over multiple repeats without confidence interval (e.g. \citep{zadghaffari2009,jahoda2007}, although insight in the variability in mixing times is a relevant characteristic.  
CFD simulations, on the other hand, are fully deterministic numerical simulations and should in principle yield a single, repeatable mixing time. This is certainly true for the Reynolds Averaged Navier Stokes (RANS) simulations which are commonly used to assess mixing behavior, and under the assumption of negligible rotor-stator interaction and macro-instability effects, can even be operated in steady-state \citep{crn2011}. But also for dynamic methods, including Large Eddy Simulations (LES) which do resolve the larger scales of the turbulent spectrum explicitly, their deterministic nature should lead to reproducibility. In practice, we did not observe this reproducibility for LES, though. Rerunning the same simulation yields a different mixing time, leading to two questions: 

\begin{enumerate}
    \item What is the origin of non-deterministic behavior in LES-mixing simulations?
    \item How does the non-deterministic behavior compare to experimental mixing time distributions?
\end{enumerate}

We conducted an initial experimental assessment combined with repeated mixing studies in two commercial CFD codes; through this short communication we share our initial observations, and want to invite the CFD community to further contribute to expanding this study. 

\section{Origin of variability in LES simulations}
Large Eddy Simulations, as any numerical simulation, are strictly deterministic. This means that if the same calculations are performed in the same manner, they should yield the same outcome. However, in practice executing simulations in exactly the same way is challenging, especially when employing parallel computing – which is essential to keep LES computationally tractable. When the order of calculations is slightly changed, this will lead to slightly different round-off of floating-point numbers, and as shown by Senoner et al. \citep{senoner2008} this will inevitably lead to ‘separation of trajectories’ in LES simulations, similar to what is observed in direct numerical simulation (DNS): after some time, exponential growth of the initial numerical round-off error leads to completely different instantaneous realizations of the flowfield. This effect could be delayed by using higher numerical precision, but eventually the same separation is observed, with the rate determined by the physical response of the system rather than numerical considerations. This distinguishes LES and DNS from RANS and laminar flow situations, where (numerical) perturbations are quenched by the diffusivity of the system. Senoner et al. do note that indeed, if the exact same simulation is conducted on the exact same system with the exact same order of calculations, LES will be fully deterministic. However, as most parallelization algorithms are based on flexible message passing, this will hardly ever be observed in practice. And even then: considering the stochastic nature of turbulence, we should wonder whether a fully determined LES simulation is truly representative of the real mixing behavior, or if it presents a particular unlikely representation of the possible realizations.

\subsection{Quantifying variability: methodology}
With the notion that LES results will inevitably be variable in practical context, we set out to quantify this variability in comparison with experimental data. Whilst more advanced experimental methods are available, we opted here for an often-used method for mixing quantification, determination of $\tau_{95}$ using a single-point conductivity measurement. We opted for this method due to (1) frequency of use, (2) ease of data processing and (3) rapid response of the conductivity probe, which in combination with an expected mixing time of ca. $25 \second$ in a $30 \liter$ bioreactor should lead to minor measurement error. 

\begin{figure}[htp]
\centering
\includegraphics[width=\columnwidth]{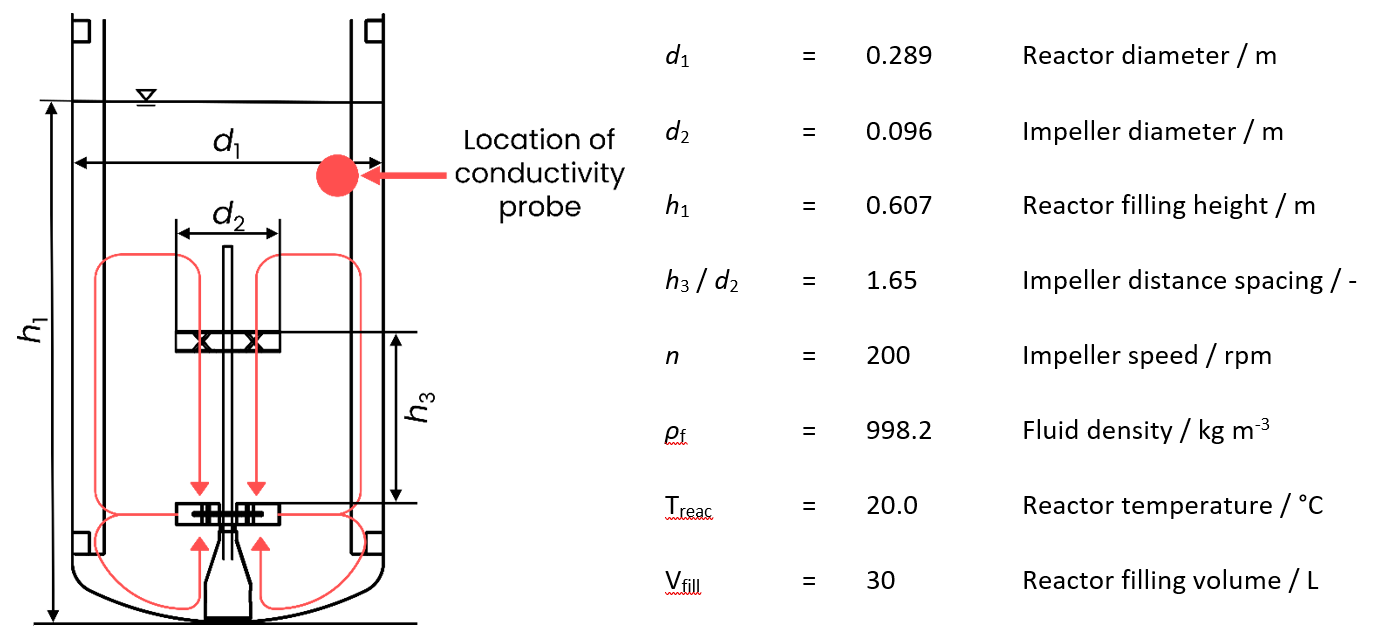}
\caption{Geometry used in the current study}
\label{fig:exp}
\end{figure}

\subsection{Experimental setup}
We conducted 11 repeats of a mixing experiment in a $30 \liter$ bioreactor with the geometry reported in figure \ref{fig:exp}, running at 200 RPM. The mixing time was determined based on $95\%$ homogenization using two WTW - IDS 
TetraCon\textsuperscript{\textregistered} 925 conductivity probes, with a response time of $1 \second$. The  probe locations and injection location are outlined in table \ref{tab:pos}.

\begin{table}

    \centering
    \caption{Location of the probes and tracer injection location used in the simulations and experiment.}

    \begin{tabular}{c c c c}
    \hline
    \textbf{Feature} & \textbf{x [m]} &  \textbf{y [m]} & \textbf{z [m]} \\
    \hline
    Probe 1 & 0.1270 & 0.4324 & 0.0180\\
    Probe 2 & 0.1270 & 0.4324 & 0.1180\\
    Injection & 0 & 0.4658 & 0.1164\\
    \hline
    \end{tabular}
    \label{tab:pos}
\end{table}

\subsection{Numerical setup}
Large Eddy Simulations were conducted using the commercial CFD packages M-Star CFD 3.12.21 and ANSYS Fluent 2024R1, which use the lattice Boltzmann (LB) and finite volume (FV) approach, respectively. In both cases the Smagorinsky subgrid model was used, with $C_S=0.1$. Due to time constraints, a mesh dependency study was not conducted in either software at this stage, as prior studies indicate mixing time determinations to be reasonably independent for this mesh resolution  \citep{haringa2018c}. 

\medskip

\textbf{M-Star Setup:} A standard D3Q19 lattice is employed with the Bhatnagar-Gross-Krook (BGK) relaxation method \citep{kuschel2021}, with explicit time integration and a courant number of Co = 0.1. A grid spacing of 0.95 mm per   point was used, yielding 90 million lattice points. A single simulation was set up with 10 different tracers, injected with a 1 second time starting at t = 25s, the first 25 seconds allows for the flowfield to establish. The simulation was repeated 5 times to yield a total of 50 mixing realizations, all simulations were conducted using an Nvidia A100 GPU and CUDA version 12.6. Mixing was quantified using  a probe consisting of a static body with nonslip conditions which represents the conductivity probe-shaft submerged in the bioreactor. A further virtual control volume below the probe-shaft is used to determine the tracer concentration. This approach closely replicates the experimental conductivity probe measurements, matching the probe-shaft diameter, length and virtual control volume to the probe control volume of the TetraCon\textsuperscript{\textregistered} 925.

\medskip

\textbf{Fluent setup:} A mesh of 3.41M elements was used, with local refinement round the impellers. The SIMPLE algorithm was set for P-V coupling, second order discretization was applied for pressure and tracer dispersion, and Bounded Central Differencing was used for Momentum, together with Second Order Implicit time integration. A sliding mesh method with a fixed timestep size of $0.01 \second$ was used. A simulation for flowfield convergence - residuals less than $O(-4)$ - was conducted using Fluent’s GPU solver on a workstation equipped with a 16 AMD Ryzen ThreadripperPRO 5955WX CPUs, 64 GB RAM, and an Nvidia GeForce RTX 3090 graphics card, storing flowfields at $1 \second$ intervals between $t = \ [25…34] \ \second$. From these files, 10 mixing simulations were constructed, each with a $1\second$ interval in tracer injection time. Each of the 10 setups was run 10 times on the DelftBlue or Snellius supercomputers, either on an Nvidia V100S or A100 GPU \citep{DHPC2024}. 

\medskip

\textbf{Setup comparison Fluent vs. M-Star:} As described above, the tracer injection procedure between M-Star and FLUENT differs somehwat. In both cases, 50 tracer injections are analyzed, but in M-Star these are based on 10 tracer injections between 5 identical simulations all started at $t=0$; as such, the 10 inter-spaced injections share a flowfield, but may be injected at different phases of a macro-instability, while the 5 repetitions are independent, meaning that macro-instabilities could originate at different places, and their impact may be different between the 5 repeats. Conversely, in Fluent, all 5 repeats start from the same starting point, which could lead to lower variability if separation of trajectories are slow compared to the simulation time. Furthermore, this means that the 10 sequential injections follow the same macro-oscillation pattern, and hence potential impact of this oscillation should be more pronounced in the 5 repeats. This difference in setup originated from convenience with the way of working in the respective software, and deserves future investigation.

\medskip
\begin{figure}[htp]
\centering
\includegraphics[width=\columnwidth]{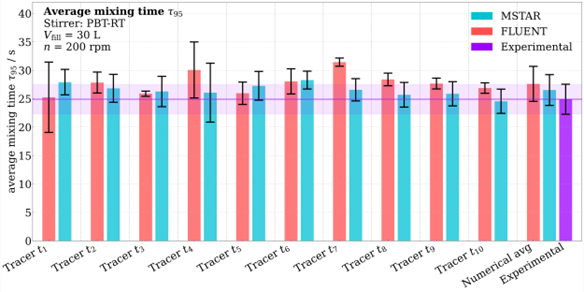}
\caption{Results of 10x 5 mixing experiments in M-Star and Fluent compared with 11 experimental repeats.  Tracers $t_1$ to $t_10$ are injected at 1 second intervals, the confidence intervals represent a single standard deviation based on 5 repeats per numerical datapoint, 11 repeats experimentally.  }
\label{fig:var}
\end{figure}

\begin{figure}[htp]
\centering
\includegraphics[width=\columnwidth]{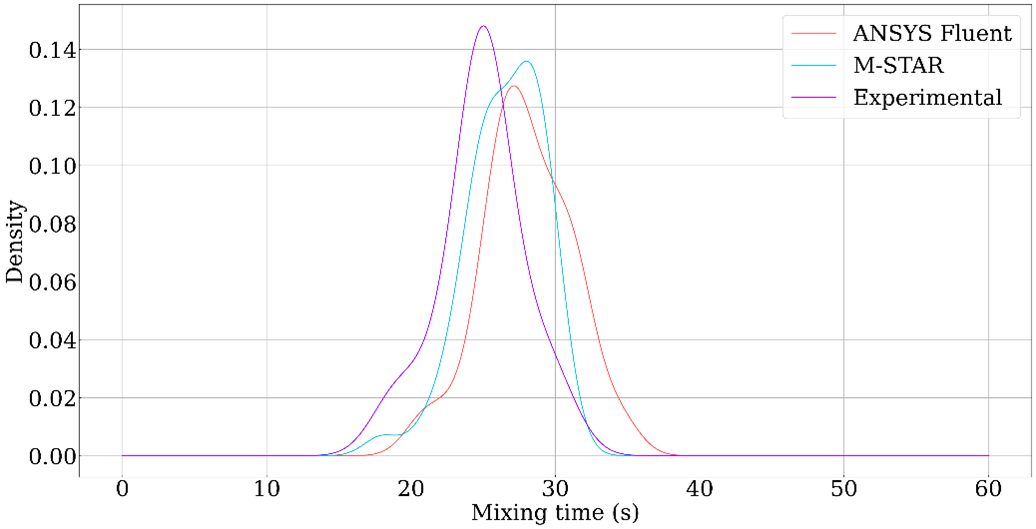}
\caption{Distributions of all mixing time experiments (experimental: 11x, M-STAR/Fluent: 50x)}
\label{fig:dist}
\end{figure}

\section{Results}

\textbf{Note:} \textit{A small offset in the angular probe position was observed between experiments and CFD. We expect this has a small impact on absolute mixing time, and no notable influence on mixing time distribution. We are currently running simulations with the correct position, and will update the figures accordingly before submission.}

\bigskip

Both experimentally and numerically a mixing time distribution was observed. Both experimentally and numerically, the mixing time was based on the maximum of the two probe readings. Figure \ref{fig:var} summarizes our observations, indicating a very good agreement in both mean mixing time and mixing time distributions between simulations and experiment; the same is observed from the overall distribution of results, not accounting for injection time (fig. \ref{fig:dist}). In M-Star, the variation between different injection times is modest; for every injection time the average lies within the experimental spread, with very similar confidence intervals. The Fluent results seem to have slightly more variation, especially confidence intervals for specific injection times. This could be related to the difference in simulation setup between the two software’s, and with that, possible differences in manifestation of macro-instabilities. However, the impact of the latter seems to be relatively small in the utilized systems; we tried to quantify the timing of tracer injections relative to the macro-instability phase using velocity probes, but did not observe strong stand-out frequencies when applying FFT on these signals, whilst such frequencies could be distinguished in e.g. 2-Rushston cases \citep{haringa2018c,haringa2018d}. Experimentally, identification of macro-instabilities in the used system is planned in future work.

\begin{figure}[htp]
\centering
\includegraphics[width=\columnwidth]{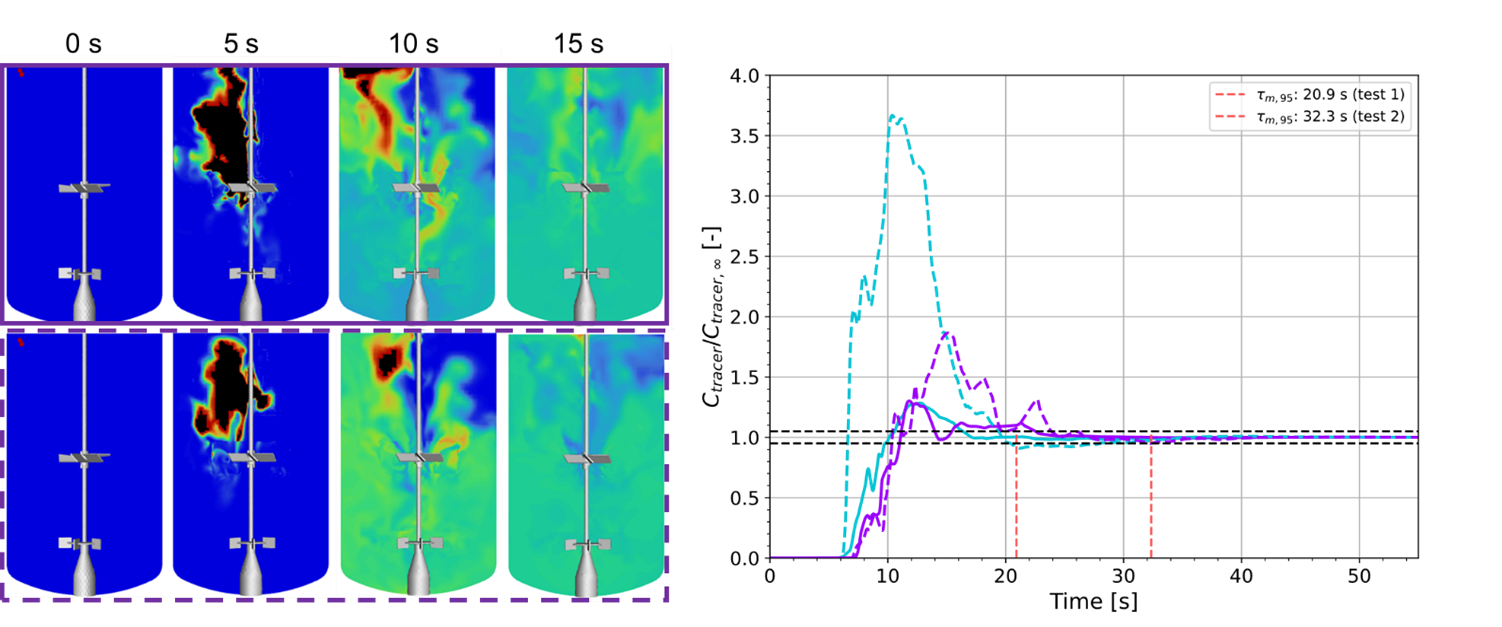}
\caption{Left: Mixing realization of 2 simulations starting from the same initial injection, at 0, 5, 10 and $15 \second$ after injection (ANSYS Fluent). Right: mixing curves (cyan: probe 1, purple: probe 2) registered by probe at  for the top (solid lines) and bottom (dashed lines) realization, and their respective 95$\%$ mixing times.}
\label{fig:trace}
\end{figure}

Figure \ref{fig:trace} illustrates the large variability observed for Fluent injection $t_1$ by highlighting two of the most distinct mixing trajectories. This strong variability contrasts our expectations; as in Fluent all repeats start from the same initial injection, we anticipated lower variability per injection time, with more variation between injection times due to synchronized macro-instabilities with all simulations originating from the same seed. These observations indicate (1) separation of trajectories occurs fast compared to overall mixing, meaning the initial state has little impact and (2) support the notion that macro-instabilities have relatively low impact with the current setup. 

\section{Conclusions and recommendations:}

Overall, we conclude that: 

\begin{enumerate}
    \item Mixing simulations in LES are practically irreproducible due to amplification of numerical errors, and as such should be reported with confidence intervals.
    \item The observed variability in repeat simulations is surprisingly well in line with experimental variability (11 repeats), implying the stochastic behavior of repeat simulations is well-captured by the filtered Navier Stokes equations.
    \item The variation between simulations is not notably sensitive to initial conditions; starting mixing simulations from the same injection state yields similar variability as starting from 0 velocity and injecting after one full mixing time.
\end{enumerate}

Our results indicate that, with the ever more common application of Large Eddy Simulations, caution should be taken in reporting simulated mixing times, and uncertainties should be reported. We do note our results are preliminary, and further investigation is recommended. In particular we suggest looking into: 

\begin{itemize}
    \item Further consideration of initial state (e.g. repeat simulations from a converged initial flowfield in M-Star and from t = 0 in Fluent).
    \item Expansion to different solvers (e.g. STAR-CCM+, OpenFOAM, etc.)
    \item Impact of resolution, turbulence model, model parameters.
    \item Impact of numerical precision: all simulations were conducted with single precision; higher precision may postpone trajectory separation and lead to lower variability. 
    \item Formal mathematical approaches to uncertainty analysis; quantification of uncertainty without requiring repeat simulations. 
    \item Impact of different hardware architecture (GPU, CPU, ...)
\end{itemize}
We invite everyone to contribute to a round-robin investigation of this issue, and would be happy to share geometry and data broadly. 

\section{Acknowledgments}
We acknowledge Jürgen Fitschen for initially pointing out this problem, Dieter Bothe, Martin Sommerfeld, Holger Marshall, Matthias Kraume and Johannes Wutz for insightful discussions.  RR is Funded by the Deutsche Forschungsgemeinschaft (DFG, German Research Foundation)—427899833. PB is supported by the Dutch National Growth Fund Cellular Agriculture. International exchange is supported by DFG CRC 1615 – 503850735. CH, HML and PB acknowledge the use of computational resources of the DelftBlue supercomputer, provided by Delft High Performance Computing Centre (https://www.tudelft.nl/dhpc), and the supercomputer Snellius (https://www.surf.nl/en/services/snellius-the-national-supercomputer). 

\bibliographystyle{elsarticle-num}
\bibliography{2025_mixrep.bib}

\end{document}